\documentclass[%
reprint,
superscriptaddress,
nofootinbib,
 amsmath,amssymb,
 aps,
]{revtex4-1}
\usepackage{graphicx}
\usepackage{subfigure}
\usepackage{dcolumn}
\usepackage{bm}
\usepackage[%
bookmarksnumbered,
bookmarksopen,
colorlinks,
citecolor=blue,
linkcolor=blue,
]{hyperref} 

\def\esym{$E_{sym}(\rho)$~}

\def\es0{$E_{sym}(\rho_0)$~}
\def\us0{$U_{sym}^{\infty}(\rho_{0})$}
\begin{document}


\title{Neutron-proton differential transverse flow in $^{132}$Sn + $^{124}$Sn collisions at 270 MeV/nucleon}

\author{Xin Huang}
\affiliation{School of Physics and Electronic Science, Guizhou Normal University, Guiyang 550025, China}
\author{Gao-Feng Wei}\email[Corresponding author. E-mail: ]{wei.gaofeng@gznu.edu.cn}
\affiliation{School of Physics and Electronic Science, Guizhou Normal University, Guiyang 550025, China}
\affiliation{Guizhou Provincial Key Laboratory of Radio Astronomy and Data Processing, Guizhou Normal University, Guiyang 550025, China}
\author{Qi-Jun Zhi}
\affiliation{School of Physics and Electronic Science, Guizhou Normal University, Guiyang 550025, China}
\affiliation{Guizhou Provincial Key Laboratory of Radio Astronomy and Data Processing, Guizhou Normal University, Guiyang 550025, China}
\author{You-Chang Yang}
\affiliation{Guizhou University of Engineering Science, Bijie 551700, China}
\affiliation{Zunyi Normal University, Zunyi 563006, China}
\author{Zheng-Wen Long}
\affiliation{College of Physics, Guizhou University, Guiyang 550025, China}


\begin{abstract}
	
Within a transport model, we study the neutron-proton differential transverse flow and its excitation function in central $^{132}$Sn + $^{124}$Sn collisions at 270 MeV/nucleon. To more accurately evaluate effects of the high-density behavior of symmetry energy \esym on this observable, we also consider the uncertainties of \esym around the saturation density $\rho_{0}$. It is shown that the neutron-proton differential transverse flow and its excitation function are mainly sensitive to the slope $L$ of \esym at $\rho_{0}$. However, the effects of low-density behavior of \esym on this observable should also be considered. Therefore, it is suggested that measurements of the neutron-proton differential transverse flow and its excitation function may provide useful complements to the constraints on $L$ extracted from the spectral pion ratio in S$\pi$RIT experiments. 

\end{abstract}

\maketitle


\section{introduction}\label{introduction}

Heavy-ion collisions (HICs) can directly generate high density nuclear matter, and thus provide the opportunity to explore the properties of strong interacting matter at extreme conditions. As an important input in simulations of HICs, the isovector component of nuclear mean field, i.e., symmetry/isovector potential, is rather uncertain because of the extreme challenge of relatively direct detection of isovector potential in experiments. Using the nucleon-nucleus scattering and ($p$,$n$) charge-exchange reactions 
~\cite{Hoff72,Pat76,Kwi78,Rap79,Jeu91,Kon03}, one can only extract limited information of isovector potential at $\rho_{0}$. 
As a result, the determination of the \esym term of the equation of state (EoS) of asymmetric nuclear matter (ANM) is still unsatisfactory compared to the relatively good determination of the isospin-independent part of the EoS of ANM~\cite{Dan02,Oert17}. Presently, the best knowledge of \esym is around $2\rho_{0}/3$, for which its value is determined to be 25.5$\pm1$ MeV from nuclear masses and isobaric analog states~\cite{Wang13,Brown13,Dan14}. At densities greater than $2\rho_{0}/3$, the uncertainties in \esym grow monotonically. For example, the \esym at $\rho_{0}$, that is commonly used as one of the criterion in fitting the parameters of the isovector potential, still has greater uncertainties than that at $2\rho_{0}/3$, e.g., 32$\pm2$ MeV~\cite{Cozma18} and 32.5$\pm3.2$ MeV~\cite{Wang18}. Also, the uncertainties for \esym at suprasaturation densities are rather larger, such as the reported value $L=106\pm37$~MeV~\cite{Bren21} 
by a calculation of $L$ correlated to the improved $R_{\rm skin}^{^{208}{\rm Pb}}$ in the PREX-II experiment~\cite{PREX-II}.

Very recently, the S$\pi$RIT collaboration reported the results of pion production in Sn + Sn collisions at 270 MeV/nucleon~\cite{Estee21,Jhang21}.
Moreover, through comparing the spectral pion ratio with the simulation from a dcQMD model~\cite{Cozma21}, they deduced that the value of $L$ is within the range from 42 to 117 MeV~\cite{Estee21}. Obviously, this value is consistent with that deduced from the correlated calculation~\cite{Bren21}.
However, the uncertainty for $L$ is still rather larger and thus need to be further constrained.
We note that the motions of energetic nucleons are directly influenced by the \esym and its $L$ value, and thus might provide a more direct detect of the \esym at suprasaturation density. This is because  these energetic participants can originate in the regions that are compressed during the violent early stages of HICs and be accelerated by the symmetry potential, resulting in
their momenta reflecting the \esym and its $L$ value.
Naturally, studies on the observables relavant to these nucleons, as an important complement to the pion spectra~\cite{Tsang17} and/or spectral pion ratio~\cite{Estee21}, may shed more lights on the \esym at suprasaturation densities. Actually, as one of this kind of observable, elliptic folw is more suitable to probe the \esym at suprasaturation density, and the corresponding studies in Au + Au collisions have already placed constraints on the \esym and its $L$ value~\cite{Cozma18,FOPI}.
As another candidate of this kind of observable, the free neutron-proton differential transverse flow has also been found to be more sensitive to the high-density behavior of
\esym\cite{LiBA00,LiBA02,Yong05,Yong06}. Therefore, we attempt to predict the sensitivities of this observable to the symmetry energy, and thus to benefit for the upcoming or ongoing measurements in S$\pi$RIT experiments. 

\section{The Model}\label{Model}

This study is carried out within an isospin- and momentum-dependent Boltzmann-Uehling-Uhlenbeck (IBUU) transport model that has incorporated constraints~\cite{Wei21b} on the momentum dependence of isovector potential from the recently reported pion data in S$\pi$RIT experiments~\cite{Jhang21}. Specifically, the nuclear interaction is an improved momentum-dependent interaction (IMDI) expressed as

\begin{eqnarray}
U(\rho,\delta ,\vec{p},\tau ) &=&A_{u}\frac{\rho _{-\tau }}{\rho _{0}}%
+A_{l}\frac{\rho _{\tau }}{\rho _{0}}+\frac{B}{2}{\big(}\frac{2\rho_{\tau} }{\rho _{0}}{\big)}^{\sigma }(1-x)  \notag \\
&+&\frac{2B}{%
	\sigma +1}{\big(}\frac{\rho}{\rho _{0}}{\big)}^{\sigma }(1+x)\frac{\rho_{-\tau}}{\rho}{\big[}1+(\sigma-1)\frac{\rho_{\tau}}{\rho}{\big]}
\notag \\
&+&\frac{2C_{l }}{\rho _{0}}\int d^{3}p^{\prime }\frac{f_{\tau }(%
	\vec{p}^{\prime })}{1+(\vec{p}-\vec{p}^{\prime })^{2}/\Lambda ^{2}}
\notag \\
&+&\frac{2C_{u }}{\rho _{0}}\int d^{3}p^{\prime }\frac{f_{-\tau }(%
	\vec{p}^{\prime })}{1+(\vec{p}-\vec{p}^{\prime })^{2}/\Lambda ^{2}},
\label{IMDIU}
\end{eqnarray}%
where $\tau=1$ for neutrons and $-1$ for protons, and $A_{u}$, $A_{l}$, $C_{u}(\equiv C_{\tau,-\tau})$ and $C_{l}(\equiv C_{\tau,\tau})$ are expressed as
\begin{eqnarray}
A_{l}&=&A_{l0} + U_{sym}^{\infty}(\rho_{0}) - \frac{2B}{\sigma+1}\notag \\
&\times&\Big{[}\frac{(1-x)}{4}\sigma(\sigma+1)-\frac{1+x}{2}\Big{]},  \\
A_{u}&=&A_{u0} - U_{sym}^{\infty}(\rho_{0}) + \frac{2B}{\sigma+1}\notag \\
&\times&\Big{[}\frac{(1-x)}{4}\sigma(\sigma+1)-\frac{1+x}{2}\Big{]},\\
C_{l}&=&C_{l0} - 2\big{(}U_{sym}^{\infty}(\rho_{0})-2z\big{)}\frac{p_{f0}^{2}}{\Lambda^{2}\ln \big{[}(4p_{f0}^{2}+\Lambda^{2})/\Lambda^{2}\big{]}},\\
C_{u}&=&C_{u0} + 2\big{(}U_{sym}^{\infty}(\rho_{0})-2z\big{)}\frac{p_{f0}^{2}}{\Lambda^{2}\ln \big{[}(4p_{f0}^{2}+\Lambda^{2})/\Lambda^{2}\big{]}}.
\end{eqnarray}
In the framework, the present IBUU model originates from the IBUU04~\cite{Das03,IBUU} and/or IBUU11~\cite{CLnote} models. However, the present version has been greatly improved to more accurate simulations of HICs as briefly discussed in the following.

First, a separate density-dependent scenario~\cite{Xu10,Chen14} for in-medium nucleon-nucleon interaction has been adopted for a more delicate treatment of the in-medium many-body force effects~\cite{Chen14}, which also affects significantly the pion production in HICs~\cite{Wei20}. Therefore, the $B$-terms in Eq.~(\ref{IMDIU}) as well in the expressions of $A_{u}$ and $A_{l}$ are different from those in Refs.~\cite{Xu15,Xu17}. 

Second, a quantity $U_{sym}^{\infty}(\rho_{0})$ proposed in Ref.~\cite{CLnote}, i.e., the value of symmetry potential at infinitely large nucleon momentum, is used to characterize the momentum dependence of the symmetry potential at $\rho_{0}$ as in Refs.~\cite{Xu15,Xu17}. It should be mentioned that this quantity is treated as a free one in Refs.~\cite{Xu15,Xu17}. However, considering that the symmetry potential with different \us0 even with the identical \esym can also lead to different pion yields in HICs, we have 
carried out a study to constrain the \us0 through comparing the pion observables of theoretical simulations for reactions $^{108}$Sn + $^{112}$Sn and $^{132}$Sn + $^{124}$Sn with the data in S$\pi$RIT experiments~\cite{Jhang21},
the central value of \us0 is constrained approximately to be $-160$~MeV, see, Ref.~\cite{Wei21b} for the details. 

\begin{table}[t]
\caption{\label{table}
The parameters $x$ and $z$ as well the corresponding $L$ values for Case-I and Case-II.}
\begin{ruledtabular}
\begin{tabular}{lccc}
\textrm{Parameters}&
\textrm{Case-I}&
\textrm{Case-II}&
\textrm{L (MeV)}\\
\colrule
$x$,~$z$ & $0.4$,~$-2.149$ & $0.491$,~$0$ & $30.3$\\
$x$,~$z$ & $0$,~$0.767$ & $-0.032$,~$0$ & $62.0$\\
$x$,~$z$ & $-0.4$,~$3.712$ & $-0.557$,~$0$ & $93.8$\\
$x$,~$z$ & $-0.8$,~$6.656$ & $-1.081$,~$0$ & $125.6$\\
\end{tabular}
\end{ruledtabular}
\end{table}
\begin{figure}[t]
	\includegraphics[width=\columnwidth]{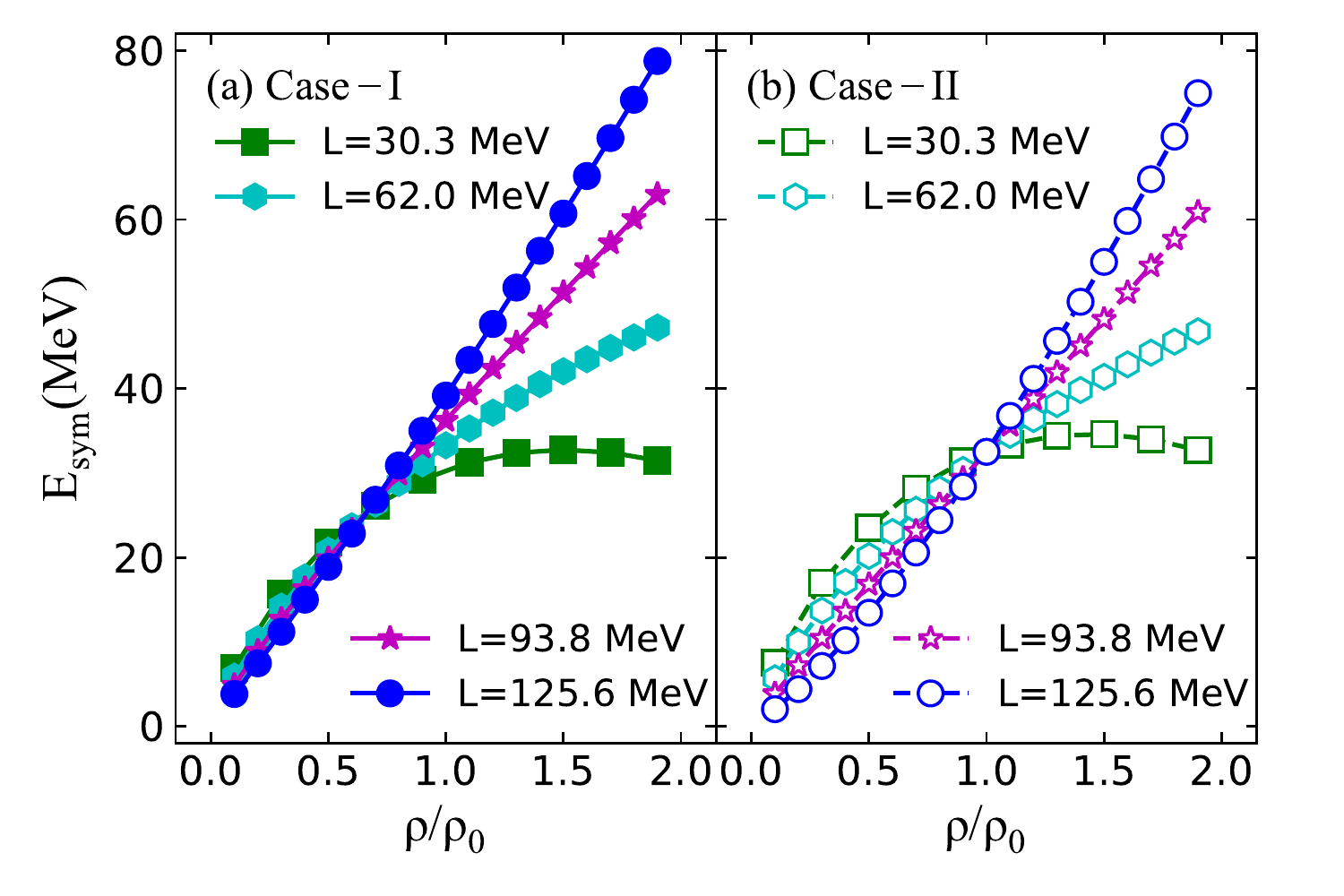}
	\caption{(Color online) Density dependence of the \esym for Case I and Case II.} \label{esym}
\end{figure}
\begin{figure}[t]
	\includegraphics[width=0.9\columnwidth]{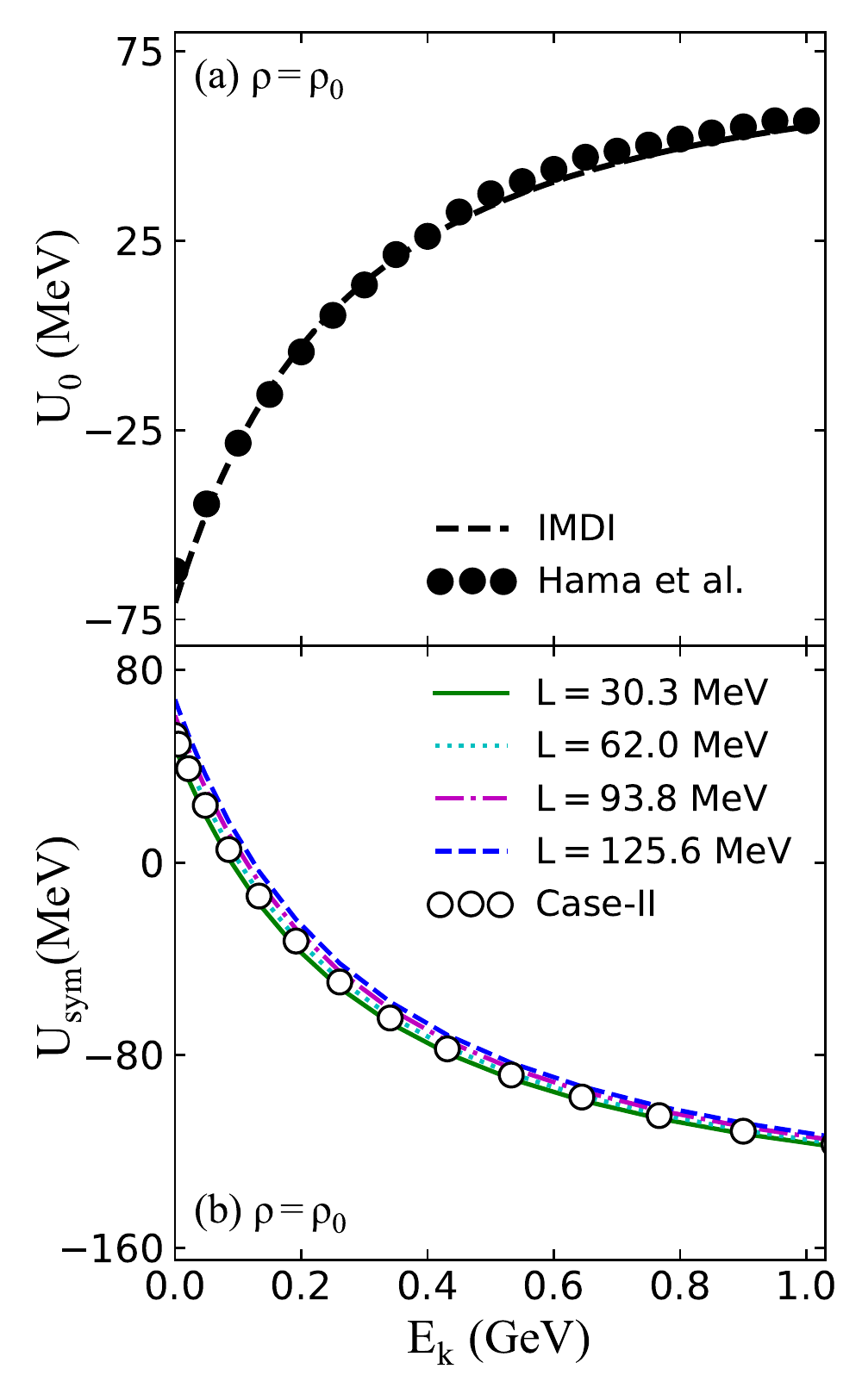}
	\caption{(Color online) Upper: Kinetic-energy dependent isoscalar potential (a) at $\rho_{0}$ in comparison with the Schr\"{o}dinger-equivalent one obtained by Hama {\it et al}. Lower: Kinetic-energy dependent isovector potential (b) at $\rho_{0}$ for Case I and Case II.}
	\label{poten}
\end{figure}

Third, considering that the \esym at $\rho_{0}$ still has greater uncertainties than that at $2\rho_{0}/3$, we therefore introduce a parameter $z$ for $C$ terms in the expressions of $C_{u}$ and $C_{l}$ as in Refs.~\cite{Xu15,Xu17} to adjust the \esym at $\rho_{0}$. As the first case, i.e., Case I, we take the value for the $z$ parameter to ensure $E_{sym}(2\rho_{0}/3)$ is $25.5$~MeV. For comparison, we also set the value of $z$ as zero and take the value for $E_{sym}(\rho_{0})$ of 32.5~MeV, as commonly used; this case is denoted as Case II. The parameters embedded in IMDI interactions are determined by fitting identical experimental/empirical constraints on properties of nuclear matter at $\rho_{0}$ for both Case I and Case II, except the value of 25.5 MeV for $E_{sym}(2\rho_{0}/3)$ used in Case I and 32.5 MeV for $E_{sym}(\rho_{0})$ used in Case II. Specifically, the values of these parameters are $A_{l0}=A_{u0}=-66.963$~MeV, $B=141.963$~MeV, $C_{l0}=-60.486$~MeV, $C_{u0}=-99.702$~MeV, $\sigma=1.2652$, $\Lambda=2.424p_{f0}$ and $U_{sym}^{\infty}(\rho_{0})=-160$~MeV, where $p_{f0}$ is the nucleon Fermi momentum in symmetry nuclear matter (SNM) at $\rho_{0}$. Moreover, to study the effects of high-density behavior of $E_{sym}(\rho)$, we adjust the $x$ parameter to obtain four different $L$ values for both Case I and Case II. The parameters $x$ and $z$ as well the corresponding $L$ values are listed in Table~\ref{table}. It is seen that except for the value of 6.656 for the $z$ parameter being slightly larger than the uncertainties of \esym at $\rho_{0}$, the values are all basically within the allowed uncertain range of \esym at $\rho_{0}$~\cite{Cozma18,Wang18}. This feature can also be seen in the density dependent \esym as shown in Fig.~\ref{esym}. It should be mentioned that except the different symmetry energy criterion the parameters for Case I and Case II can lead to identical properties for nuclear matter at $\rho_{0}=0.16$~fm$^{-3}$, i.e., the binding energy $-16$~MeV, the incompressibility $K_{0}=230$~MeV for SNM, the isoscalar effective mass $m^{*}_{s}=0.7m$, the isoscalar potential at infinitely large nucleon momentum $U^{\infty}_{0}(\rho_{0})=75$~MeV as well $U^{\infty}_{sym}(\rho_{0})=-160$~MeV. Shown in Fig.~\ref{poten} are the isoscalar and isovector potentials at $\rho_{0}$ for Case I and Case II. Since the parameters $x$ and $z$ only affect the isovector properties of ANM, the isoscalar potentials for Case I and Case II are the same and also compatible with the results of Hama {\it et al}~\cite{Hama90,Buss12}. Moreover, since the $x$ parameter only affects the isovector properties of ANM at nonsaturation densities for both Case I and Case II, as well the identical $z$ values used for Case II, we can observe that the isovector potentials at $\rho_{0}$ even with different $L$ values for Case II are also identical to each other. In contrast, since the $z$ parameter is used to ensure $E_{sym}(2\rho_{0}/3)$ to be $25.5$~MeV for Case I by adjusting $E_{sym}(\rho_{0})$, it naturally affects isovector potentials at $\rho_{0}$. Therefore, we can see that the isovector potentials at $\rho_{0}$ with different $L$ for Case I are slightly different due to the different $z$ values used for different $L$ in Case I. However, the differences of isovector potentials in Case I from those in Case II are very tiny. 

Fourth, to more accurately simulate HICs, we also give detailed consideration of the $\Delta$ and pion potentials as well as electromagnetic effects in HICs, see, Refs.~\cite{Wei18a,Wei18b,Wei21,Wei21b} for more details. 

\section{Results and Discussions}\label{Results and Discussions}

\begin{figure}[t]
	\includegraphics[width=\columnwidth]{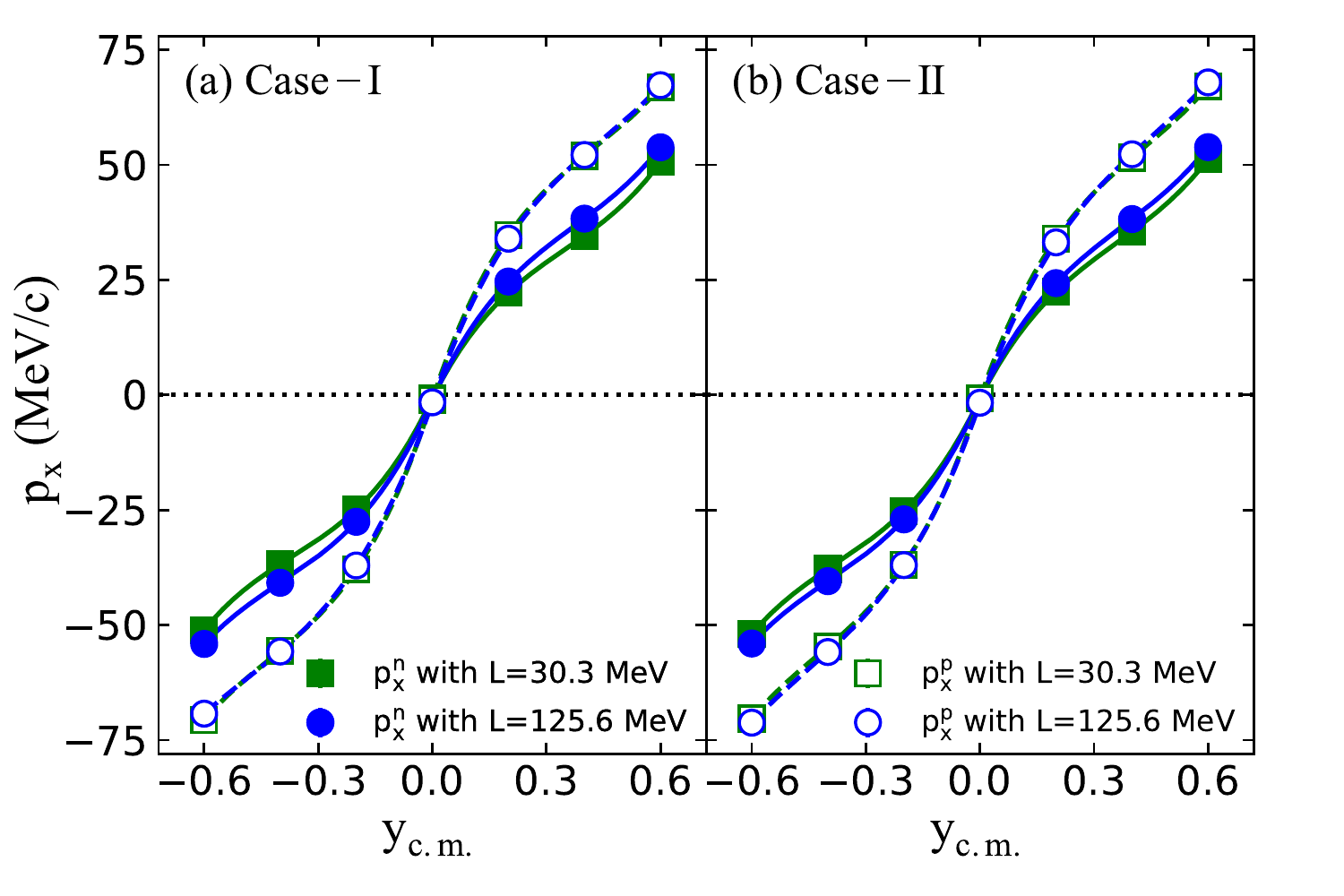}
	\caption{(Color online) Differential transverse flows of free neutrons $p_{x}^{n}$ and protons $p_{x}^{p}$ as a function of the center-of-mass rapidity $y_{c.m.}$.} \label{px}
\end{figure}
\begin{figure}[t]
	\includegraphics[width=\columnwidth]{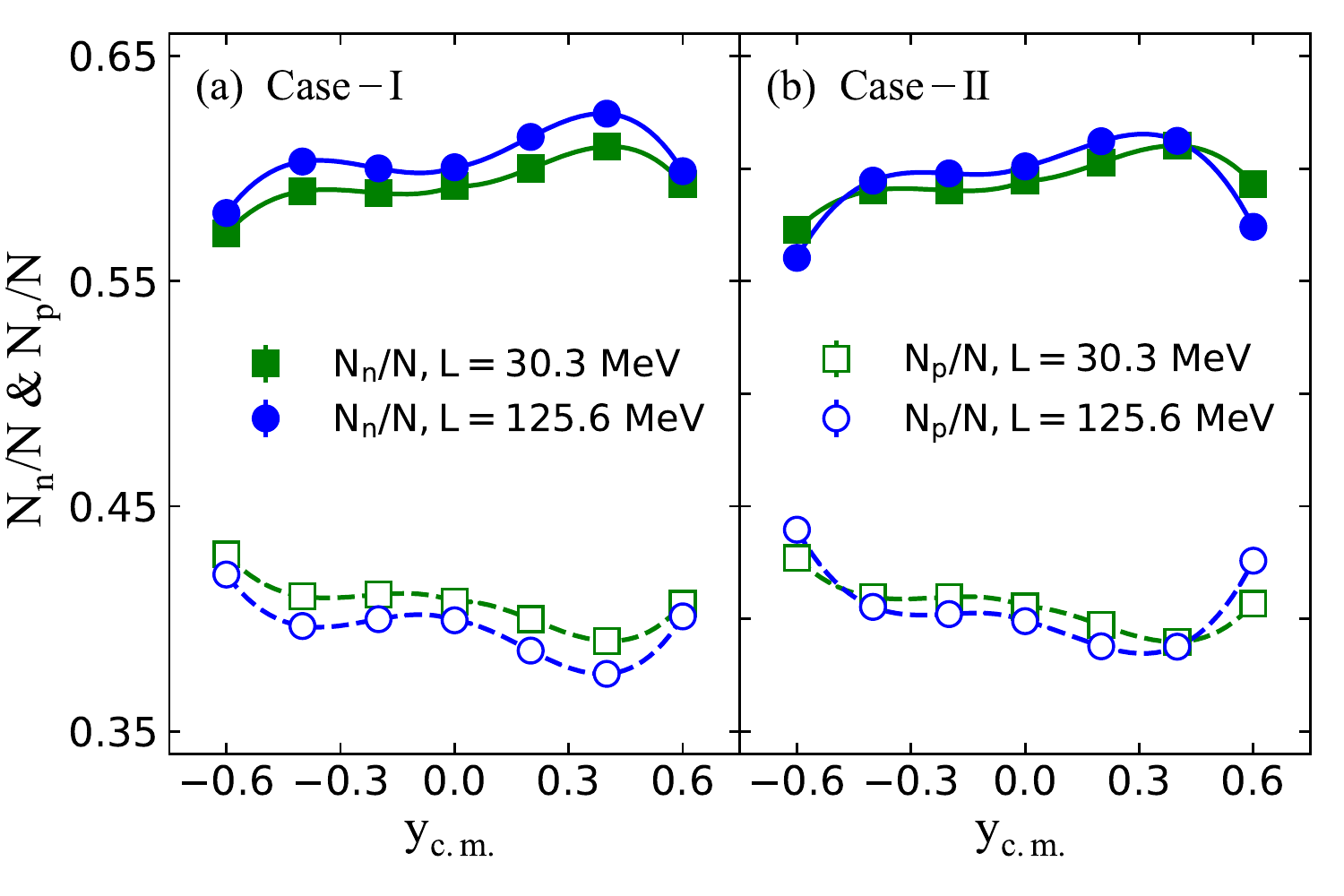}
	\caption{(Color online) $N_{n}(y)/N(y)$ and $N_{p}(y)/N(y)$ with $L=30.3$~MeV and $125.6$~MeV as a function of the center-of-mass rapidity $y_{c.m.}$.} \label{isofrac}
\end{figure}
\begin{figure}[t]
	\includegraphics[width=\columnwidth]{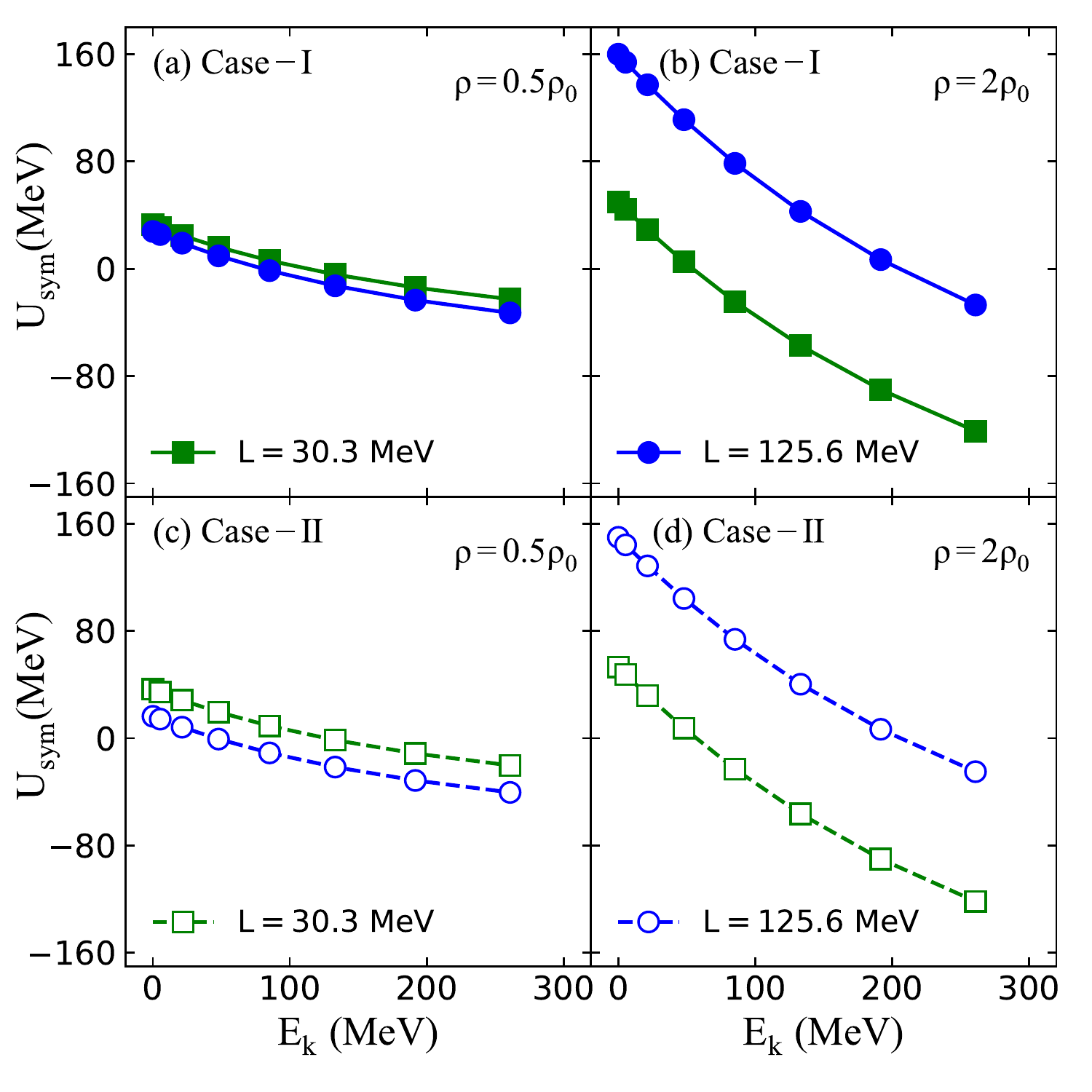}
	\caption{(Color online) Kinetic-energy dependent isovector potential at $\rho=0.5\rho_{0}$ and $\rho=2\rho_{0}$ for Case I and Case II.} \label{sympot}
\end{figure}

Now, we present the results for the $^{132}$Sn + $^{124}$Sn reaction at 270 MeV/nucleon with an impact parameter of $b=3$ fm. 
The free neutron-proton differential transverse flow is measured by~\cite{LiBA00,LiBA02,Yong05,Yong06}
\begin{eqnarray}
p_{x}^{np}(y)&=&\frac{1}{N(y)}\sum_{i=1}^{N(y)}p_{x_{i}}\tau_{i}\notag\\
&=&\frac{N_{n}(y)}{N(y)}<p_{x}^{n}(y)>-\frac{N_{p}(y)}{N(y)}<p_{x}^{p}(y)>,
\end{eqnarray}
where $N_{n}(y)$, $N_{p}(y)$ and $N(y)$ denote, respectively, the number of free neutrons, protons, and total nucleons with local densities less than $\rho_{0}/8$ at rapidities $y$, and $\tau_{i}$ is $1$ for neutrons and $-1$ for protons. From this formula, we can see that the isospin fractionation effects are incorporated into the collective flow~\cite{Dan85} of both neutrons and protons through $N_{n}(y)/N(y)$ and $N_{p}(y)/N(y)$. To understand this effect, we show first in Fig.~\ref{px} the collective flow of free neutrons and protons as a function of the center-of-mass rapidity. First, it is seen that, consistent with the previous result in Ref.~\cite{LiBA00}, the proton flow is less sensitive to $L$ but higher than the neutron flow due to the Coulomb repulsive effects. Second, with a certain $L$, we can observe that the transverse flows either for neutrons or protons are not changed essentially in the left and righ panels of Fig.~\ref{px}. This means that the transverse flow is relatively insensitive to the actual value for \esym at any given density, but is mainly sensitive to the slope of \esym, which governs the pressure that the symmetry energy provides in a neutron-rich system as well dense astrophysical environments such as neutron stars. It should be mentioned that this finding is similar to that observed in the ratios/differences of neutron vs proton elliptic flows in Refs.~\cite{Cozma18,Cozma11} as well that observed much earlier in Ref.~\cite{Dan02}.
Nevertheless, the sensitivities of transverse flows to $L$ are not obvious either for neutrons or protons. Shown in Fig.~\ref{isofrac} are  $N_{n}(y)/N(y)$ and $N_{p}(y)/N(y)$ as a function of the center-of-mass rapidity, respectively. First, since the mass between the target and projectile has a little difference, we thus observe that the shapes of both $N_{n}(y)/N(y)$ and $N_{p}(y)/N(y)$ are a little asymmetric between the target and projectile rapidities. Second, it is seen that 
varying $L$ from $30.3$ to $125.6$~MeV causes more (less) free neutrons (protons) on the whole for both Case I and Case II. Certainly, we can also find that effects of $L$ on either $N_{n}(y)/N(y)$ or $N_{p}(y)/N(y)$ are reduced somewhat in mid-rapidities for Case II compared to those for Case I. Also, the effects even reverse at the target and/or projectile rapidities. To understand these observations, we show in Fig.~\ref{sympot} the isovector potential of Case I and Case II at both $0.5\rho_{0}$ and $2\rho_{0}$. First, it is seen that $L$ affects the isovector potentials mainly at high densities, and thus the high-density behavior of \esym dominates $N_{n}(y)/N(y)$ and $N_{p}(y)/N(y)$. Moreover, varying $L$ from $30.3$ to $125.6$~MeV, the neutrons feel stronger repulisve effects while the protons feel stronger attractive effects on the whole for both Case I and Case II. Naturally, we can observe more free neutrons but less free protons with larger values for $L$. Second, we observe that the differences of isovector potentials at $2\rho_{0}$ between Case I and Case II are negligible, while the differences of isovector potentials at $0.5\rho_{0}$ with different $L$ are smaller in Case I than those in Case II. This is exactly due to the more accurate criterion, i.e., $E_{sym}(2\rho_{0}/3)=25.5$~MeV, used in Case I that reduces the effects of $L$ on the isovector potentials and $E_{sym}(\rho)$ at subsaturation densities. Moreover, since the effects of $L$ on the isovector potentials at low densities are opposite to those at high densities, the effects of high-density symmetry energy/potential on observables at compress stages are likely to be smeared out by low density symmetry energy/potential at the expansion stages. In particular, nucleons with target and/or projectile rapidities experience longer time in the expansion stages and emit later compared to those at midrapidities, the effects of low density symmetry energy/potential on observables are naturally considerable. This is the reason we observe in Case II that the isospin fractionation effects are reduced slightly at midrapidities but reversed at target and/or projectile rapidities.

\begin{figure}[t]
	\includegraphics[width=\columnwidth]{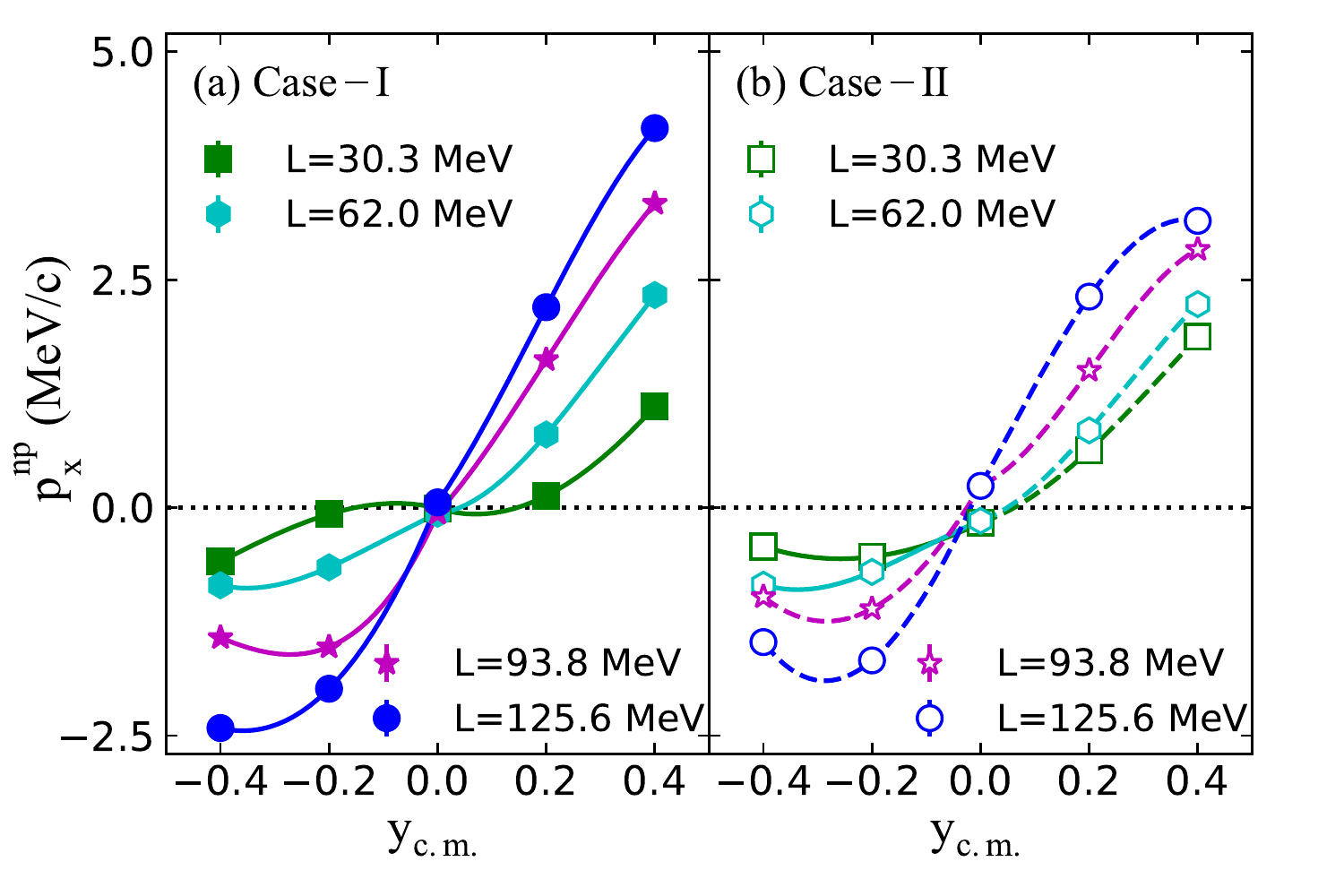}
	\caption{(Color online) Neutron-proton differential transverse flows as a function of the center-of-mass rapidity $y_{c.m.}$.} \label{pxnp}
\end{figure}

\begin{figure}[t]
	\includegraphics[width=\columnwidth]{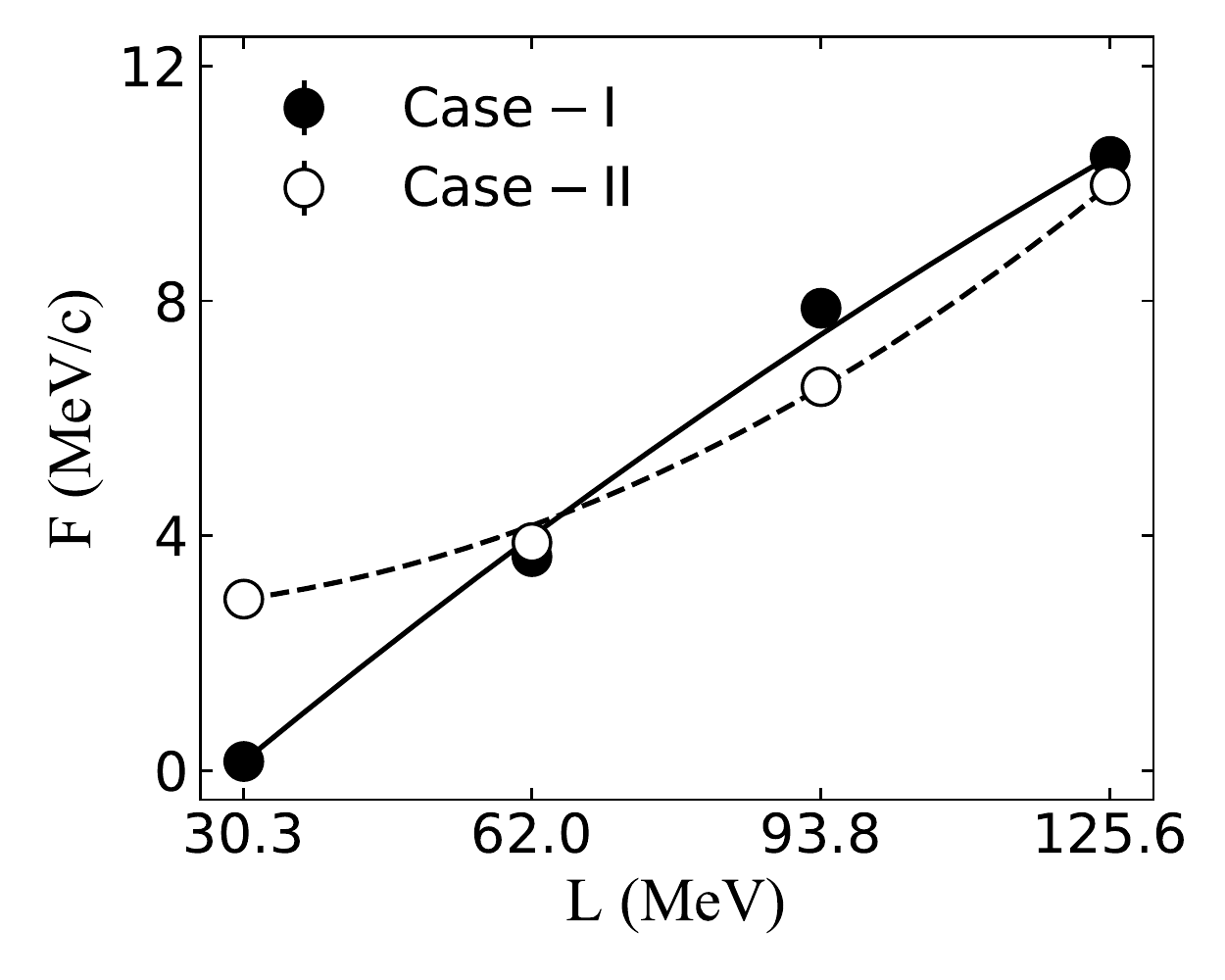}
	\caption{Excitation function of the neutron-proton differential transverse flows at the mid-rapidity $|y_{c.m.}|\le{0.1}$ as a function of $L$. The lines are drawn to guide the eye.} \label{slope}
\end{figure}

Now, we turn to the free neutron-proton differential transverse flows in the same reaction. Shown in Fig.~\ref{pxnp} are the free neutron-proton differential transverse flows as a function of the center-of-mass rapidity. First, it is observed that the free neutron-proton differential transverse flow inherits the asymmetry in shapes from  $N_{n}(y)/N(y)$ and $N_{p}(y)/N(y)$.
Second, it is seen that this observable indeed combines constructively the in-plane transverse momenta generated by the isovector potentials while reducing significantly influences of the isoscalar potentials of both neutrons and protons as indicated in Ref.~\cite{LiBA02}, and thus it is more sensitive to $L$. Moreover, comparing observations between Case I and Case II, we can find that the sensitivities of this observable to $L$ are indeed reduced in Case II. Therefore, to show clearly this observation and facilitate the experimental measurement, we calculate the excitation function of this observable, which is defined as~\cite{LiBA02b}
\begin{equation}\label{excitation-func}
F={\Big(} \frac{d<p_{x}^{np}>}{dy_{c.m.}} {\Big)}_{y_{c.m.}=0}.
\end{equation}
Shown in Fig.~\ref{slope} is the excitation function of neutron-proton differential transverse flows as a function of $L$. It is seen that effects of $L$ on the excitation function are rather obvious for both Case I or Case II, and thus can be used to further verify the extracted $L$ from the spectral pion ratio in S$\pi$RIT experiments~\cite{Estee21}.
Nevertheless, we can also find that the curve of excitation function in Case I indeed is more steeper compared to that in Case II due to fewer effects of low-density $E_{sym}(\rho)$. This feature indicates that the uncertainties of \esym around the $\rho_{0}$ should also be considered when using this observable to probe the high-density behavior of $E_{sym}(\rho)$. Actually, the authors of Ref.~\cite{Stone17} have already discussed systematically the effects of uncertainties of low density $E_{sym}(\rho)$ on the determination of high-density $E_{sym}(\rho)$.

Before ending this part, we give two useful remarks. First, the value of $-160$~MeV used for \us0 in this study is more negative than that used in other models, e.g., $-115$~MeV, used in Refs.~\cite{Xu15,Xu17}. This large and negative \us0 appears to have the feature that interactions between two protons or two neutrons with relatively large velocity will be far more attractive than those between a neutron and proton moving at the same relative velocity. The reasons might be due to different isospin states in the former and latter that may originates from the different constituent quarks between neutrons and protons. This issue also deserves serious consideration. Second, we predict the emission of nucleons as free particles and do not consider the clustering effects of nucleons that may change the quantitative results of the present study. Therefore, it will be interesting to see how clusters change the quantitative results of the present study.

\section{Summary}\label{Summary}
In conclusion, we have studied the free neutron-proton differential transverse flow and its excitation function in $^{132}$Sn + $^{124}$Sn collisions at 270 MeV/nucleon within a transport model. It is found that the sensitivities of free neutron-proton differential transverse flow and its excitation function to $L$ are rather obvious, and thus can be used to further verify the extracted $L$ from the spectral pion ratio in S$\pi$RIT experiments. Therefore, we conclude that measurements of the neutron-proton differential transverse flow and its excitation function may provide useful complements to the constraints on $L$ extracted from measurements of the spectral pion ratio in S$\pi$RIT experiments. Moreover, after examining the effects of \esym around $\rho_{0}$ on this observable, 
it is also suggested that the uncertainties of \esym around $\rho_{0}$ should also be considered when using the free neutron-proton differential transverse flow to probe the high-density behavior of $E_{sym}(\rho)$.

\begin{acknowledgments}
G.-F. Wei would like to thank Profs. Wei Zuo and Bo-Chao Liu for helpful discussions.
This work is supported by the National Natural Science Foundation of China under grant Nos.11965008, 11405128, 11865019 and Guizhou Provincial Science and Technology Foundation under Grant No.[2020]1Y034, and the PhD-funded project of Guizhou Normal university (Grant No.GZNUD[2018]11).
\end{acknowledgments}


\begin{thebibliography}{99}
	
\bibitem{Hoff72} G. W. Hoffmann, and W. R. Coker, Phys. Rev. Lett. \textbf{29}, 227 (1972).

\bibitem{Pat76} D. M. Patterson {\it et al}., Nucl. Phys. A \textbf{263}, 261 (1976).

\bibitem{Kwi78} K. Kwiatkowski {\it et al}., Nucl. Phys. A \textbf{301}, 349 (1978).

\bibitem{Rap79} J. Rapaport {\it et al}., Nucl. Phys. A \textbf{330}, 15 (1979).

\bibitem{Jeu91} J. P. Jeukenne, C. Mahaux, and R. Sartor, Phys. Rev. C \textbf{43}, 2211 (1991).

\bibitem{Kon03} A. J. Koning {\it et al}., Nucl. Phys. A \textbf{713}, 231 (2003).

\bibitem{Dan02} P. Danielewicz, R. Lacey, and W. G. Lynch, Science \textbf{298}, 1592 (2002).

\bibitem{Oert17} M. Oertel, M. Hempel, T. KI\"{a}hn, and S. Typel, Rev. Mod. Phys. \textbf{89}, 015007 (2017).

\bibitem{Wang13} N. Wang, L. Ou, and M. Liu, Phys. Rev. C \textbf{87}, 034327 (2013).

\bibitem{Brown13}B. A. Brown, Phys. Rev. Lett. \textbf{111}, 232502 (2013).

\bibitem{Dan14}P. Danielewicz, J. Lee, Nucl. Phys. A \textbf{922}, 1 (2014).

\bibitem{Cozma18}M. D. Cozma, Eur. Phys. J. A \textbf{54}, 40 (2018).

\bibitem{Wang18}R. Wang, L. W. Chen, and Y. Zhou, Phys. Rev. C \textbf{98}, 054618 (2018).

\bibitem{Bren21} B. T. Reed, F. J. Fattoyev, C. J. Horowitz, and J. Piekarewicz, Phys. Rev. Lett. \textbf{126}, 172503 (2021).

\bibitem{PREX-II} D. Adhikari {\it et al}. (PREX Collaboration), Phys. Rev. Lett. \textbf{126}, 172502 (2021).  

\bibitem{Jhang21}G. Jhang, J. Estee, J. Barney, G. Cerizza, M. Kaneko, J. W. Lee, W. G. Lynch, T. Isobe {\it et al}.
(S$\pi$RIT Collaboration),
M. Colonna, D. Cozma, P. Danielewicz, H. Elfner, N. Ikeno, C. M. Ko, J. Mohs, D. Oliinychenko {\it et al}.
(TMEP Collaboration), Phys. Lett. \textbf{B} 813, 136016 (2021).

\bibitem{Estee21} J. Estee, W. G. Lynch, C. Y. Tsang, J. Barney, G. Jhang, M. B. Tsang, R. Wang, M. Kaneko {\it et al}. (S$\pi$RIT Collaboration), and M. D. Cozma, Phys. Rev. Lett. \textbf{126}, 162701 (2021).

\bibitem{Cozma21} M. D. Cozma, and M. B. Tsang, Eur. Phys. J. A \textbf{57}, 309 (2021).


\bibitem{Tsang17}M. B. Tsang, J. Estee, H. Setiawan, W. G. Lynch, J. Barnery, M. B. Chen, G. Cerizza, P. Danielewicz {\it et al}., Phys. Rev. C \textbf{95}, 044614 (2017). 

\bibitem{FOPI} W. Reisdorf {\it et al.} (FOPI Collaboration), Nucl. Phys. A \textbf{781}, 459 {2007}; Nucl. Phys. A \textbf{848}, 366 (2010); Nucl. Phys. A \textbf{876}, 1 (2012).

\bibitem{LiBA00} B. A. Li, Phys. Rev. Lett. \textbf{85}, 4221 (2000).

\bibitem{LiBA02} B. A. Li, Phys. Rev. Lett. \textbf{88}, 192701 (2002).

\bibitem{Yong05}G. C. Yong, B. A. Li, and W. Zuo, Chin. Phys. Lett. \textbf{22}, 2226 (2005).

\bibitem{Yong06}G. C. Yong, B. A. Li, and L. W. Chen, Phys. Rev. C \textbf{74}, 064617 (2006).

\bibitem{Wei21b} G. F. Wei, X. Huang, Q. J. Zhi, A. J. Dong, C. G. Peng, and Z. W. Long, arXiv: 2112.13518.

\bibitem{Das03} C. B. Das, S. Das Gupta, C. Gale, and B. A. Li, Phys. Rev. C \textbf{67}, 034611 (2003).

\bibitem{IBUU}B. A. Li, C. B. Das, S. Das Gupta, and C. Gale, Phys. Rev. C \textbf{69}, 011603(R) (2004).

\bibitem{CLnote} L. W. Chen, B. A. Li, A note of an improved MDI interaction for transport model simulations of heavy ion collisions , Texas A\&M University-Commerce, 2010 (Unpublished).

\bibitem{Xu10}C. Xu, B. A. Li, Phys. Rev. C \textbf{81}, 044603 (2010).

\bibitem{Chen14}L. W. Chen, C. M. Ko, B. A. Li, C. Xu, and J. Xu, Eur. Phys. J. A \textbf{50}, 29 (2014).

\bibitem{Wei20} G. F. Wei, C. Xu, W. Xie, Q. J. Zhi, S. G. Chen, and Z. W. Long, Phys. Rev. C \textbf{102}, 024614 (2020).

\bibitem{Xu15} J. Xu, L. W. Chen, and B. A. Li, Phys. Rev. C \textbf{91}, 014611 (2015).

\bibitem{Xu17} H. Y. Kong, J. Xu, L. W. Chen, B. A. Li, and Y. G. Ma, Phys. Rev. C \textbf{95}, 034324 (2017).

\bibitem{Hama90} S. Hama, B. C. Clark, E. D. Cooper, H. S. Sherif, and R. L. Mercer, Phys. Rev. C \textbf{41}, 2737 (1990).

\bibitem{Buss12} O. Buss, T. Gaitanos, K. Gallmeister, H. van Hees, M. Kaskulov, O. Lalakulich, A. B. Larionoov, T. Leitner, J. Weil, and U. Mosel, Phys. Rep. \textbf{512}, 1 (2012).

\bibitem{Wei21} G. F. Wei, C. Liu, X. W. Cao, Q. J. Zhi, W. J. Xiao, C. Y. Long, and Z. W. Long, Phys. Rev. C \textbf{103}, 054607 (2021).

\bibitem{Wei18a}G. F. Wei, B. A. Li, G. C. Yong, L. Ou, X. W. Cao, and X. Y. Liu, Phys. Rev. C \textbf{97}, 034620 (2018).

\bibitem{Wei18b}G. F. Wei, G. C. Yong, L. Ou, Q. J. Zhi, Z. W. Long, and X. H. Zhou, Phys. Rev. C \textbf{98}, 024618 (2018).

\bibitem{Dan85} P. Danielewicz, and G. Odyniec, Phys. Lett. B \textbf{157}, 146 (1985).

\bibitem{Cozma11}M. D. Cozma, Phys. Lett. B \textbf{700}, 139 (2011).

\bibitem{LiBA02b} B. A. Li, Nucl. Phys. A \textbf{708}, 365 (2002).

\bibitem{Stone17}J. R. Stone, P. Danielewicz, and Y. Iwata, Phys. Rev. C \textbf{96}, 014612 (2017).



	
\end{thebibliography}
\end{document}